\documentclass{fmj2010}
\def\fgrst{\textit{Fermi Gamma-Ray Space Telescope}}
\def\fermilat{\textit{Fermi}/LAT}

\begin{document}
   \title{Physical properties of blazar jets from VLBI observations}

   \author{Andrei Lobanov}

   \institute{Max-Planck-Institut f\"ur Radioastronomie, Auf dem H\"ugel 69, 53121 Bonn, Germany}

   \abstract{Relativistic jets, formed in the vicinity of central
supermassive black holes in AGN, show ample evidence connecting them
to physical conditions in the accretion disc and broad-line region.
The jets are responsible for a large fraction of
non-thermal continuum emission (particularly during powerful flares),
which makes understanding their physics an important aspect of studies
of blazars characterised by profound flaring activity arising from
extremely compact regions. 
Imaging and polarimetry of radio emission on milliarcsecond scales
provided by very long baseline interferometry (VLBI) offers a range of
possibilities for studying ultra-compact regions in relativistic jets
and relating them to main manifestations of the blazar activity in
AGN.  Simultaneous monitoring of optical/high energy variability and
evolution of parsec-scale radio structures yields arguably the most
detailed picture of the relation between acceleration and propagation
of relativistic flows and non-thermal continuum generation in blazars.
Opacity effects provide a measure of magnetic
field strength on scales down to $\sim 1000$ gravitational radii and
trace the distribution of broad-line emitting material. 
Correlations observed between parsec-scale radio emission and
optical and gamma-ray continuum
indicate that a significant fraction of non-thermal continuum may be
produced (particularly during flares) in extended regions of
relativistic jet at distances up to 10 parsecs from the central
engine. Combined with studies of jet component ejections and X-ray
variability, these correlations also suggest that time delays, nuclear
opacity, and jet acceleration may have a pronounced effect on the
observed broad-band variability and instantaneous spectral energy
distribution (SED). These effects will be reviewed below and discussed in
the context of deriving accurate and self-consistent models for
central regions of blazars.  
}

   \maketitle
%

\section{Introduction}

In the current astrophysical paradigm for active galactic nuclei
(AGN), each constituent of an AGN contributes to a specific domain in
the broad-band spectral energy distribution (Ghisellini \& Tavecchio
\cite{lobanov_ghisellini2009}).  Variability of continuum flux in AGN,
signalling the activity of the central engine, is detected throughout
the entire electromagnetic spectrum, on time-scales from days to
years.  Substantial progress achieved during the past decade in
studies of active galactic nuclei (see Lobanov \& Zensus
\cite{lobanov_lobanov2006} for a review of recent results) has brought
an increasingly wider recognition of the ubiquity of relativistic
outflows (jets) in AGN (Falcke \cite{lobanov_falcke2001}, Zensus
\cite{lobanov_zensus1997}) and in blazars in particular.

Understanding the physics of blazars jets has acquired a particular
importance after the launch of \fgrst, as the compact, relativistic
flows contribute strongly to the broad-band continuum -- a fact that
is still somewhat overlooked in physical models describing high-energy
emission from AGN.

Emission properties, dynamics, and evolution of an extragalactic jet
are intimately connected to the characteristics of the supermassive
black hole, accretion disk and broad-line region in the nucleus of the
host galaxy (Lobanov \cite{lobanov_lobanov2008}).  The jet
continuum emission is dominated by non-thermal synchrotron and
inverse-Compton radiation (Unwin et al. \cite{lobanov_unwin1997}). The
synchrotron mechanism plays a more prominent role in the radio domain,
and the properties of the emitting material can be assessed using the
turnover point in the synchrotron spectrum (Lobanov
\cite{lobanov_lobanov1998b}, Lobanov \& Zensus
\cite{lobanov_lobanov1999}), synchrotron self-absorption (Lobanov
\cite{lobanov_lobanov1998a}), and free-free absorption in the ambient plasma (Walker
et al. \cite{lobanov_walker2000}, Kadler et
al. \cite{lobanov_kadler2004}).

There is growing evidence for relativistic flows contributing
substantially to generation of non-thermal continuum in the optical
(Arshakian et al. \cite{lobanov_arshakian2010}, Le\'on-Tavares et
al. \cite{lobanov_leon2010}, Jorstad et
al. \cite{lobanov_jorstad2010}), X-ray (Unwin et
al. \cite{lobanov_unwin1997}, D'Arcangelo et
al. \cite{lobanov_darcangelo2007}, Marscher et
al. \cite{lobanov_marscher2008}, Soldi et
al. \cite{lobanov_soldi2008}, Chatterjee et al. \cite{lobanov_chatterjee2009}), $\gamma$-ray (Otterbein et
al. \cite{lobanov_otterbein1998}, Jorstad et
al. \cite{lobanov_jorstad2010}, Marscher et
al. \cite{lobanov_marscher2010}, Savolainen et
al. \cite{lobanov_savolainen2010}, Schinzel et al. \cite{lobanov_schinzel2010}) and TeV (Piner \& Edwards
\cite{lobanov_piner2004}, Charlot et al. \cite{lobanov_charlot2006},
Acciari et al. \cite{lobanov_acciari2009}) domains.  Accurate spatial
localisation of production sites of variable non-thermal continuum
emission in AGN is therefore instrumental for understanding the
mechanism for release and transport of energy in active galaxies.

In the radio regime, very long baseline interferometry (VLBI) enables
direct imaging of spatial scales comparable the gravitational radius,
$R_\mathrm{g} = G\,M_\mathrm{bh}/c^2$, of the central black hole in
AGN using ground VLBI observations at 86\,GHz and higher ({\em cf.,}
GMVA\footnote{Global Millimeter VLBI Array;
http://www.mpifr-bonn.mpg.de/div/vlbi/globalmm} observations;
Krichbaum et al. \cite{lobanov_krichbaum2008}) and space VLBI
observations at centimetre wavelengths (Takahashi et
al. \cite{lobanov_takahashi2009}). Such high-resolution radio
observations also access directly the regions where the jets are
formed (Junor et al. \cite{lobanov_junor1999}), and trace their
evolution and interaction with the nuclear environment (Lobanov
\cite{lobanov_lobanov2007,lobanov_lobanov2008} and Middelberg \& Bach
\cite{lobanov_middelberg2008}).  Evolution of compact radio emission
from several hundreds of extragalactic relativistic jets is now
systematically studied with dedicated monitoring programs and large
surveys using very long baseline interferometry (such as the 15\,GHz
VLBA\footnote{Very Long Baseline Array of National Radio Astronomy
Observatory, USA} survey (Kellermann et
al. \cite{lobanov_kellermann2004}) the
MOJAVE\footnote{http://www.physics.purdue.edu/MOJAVE} survey (Lister
\& Homan \cite{lobanov_lister2005}) and a dedicated 22/43/86\,GHz VLBA
gamma-ray blazar monitoring program at the Boston
University\footnote{http://www.bu.edu/blazars} (Jorstad et
al. \cite{lobanov_jorstad2001}).  These studies, combined with optical
and X-ray studies, yield arguably the most detailed picture of the
vicinity of supermassive black holes in AGN (Marscher
\cite{lobanov_marscher2005}).

Presented below is a brief (and certainly incomplete) summary of
recent results from VLBI studies of compact extragalactic radio
sources, outlining the physical properties of relativistic
parsec-scale jets and their relation to supermassive black holes,
accretion disks and broad-line regions in prominent blazars.

\section{Physics of compact jets}

Jets in active galaxies are formed in the immediate vicinity of the
central black hole (Camenzind \cite{lobanov_camenzind2005}), at
distances of $10$--$10^2\,R_\mathrm{g}$ (Meier
\cite{lobanov_meier2009}). The jets carry away a fraction of the
angular momentum and energy stored in the accretion flow (Blandford \&
Payne \cite{lobanov_blandford1982}, Hujeirat et
al. \cite{lobanov_hujeirat2003} or corona (in low luminosity AGN;
Merloni \& Fabian \cite{lobanov_merloni2002}) and in the rotation of
the central black hole (Blandford \& Znajek 1977, Koide et
al. \cite{lobanov_koide2002}, Komissarov
\cite{lobanov_komissarov2005}, Semenov et
al. \cite{lobanov_semenov2004}).

The production of highly-relativistic outflows requires a large
fraction of the energy to be converted to Poynting flux in the very
central region (Sikora et al. \cite{lobanov_sikora2005}).  The
efficiency of this process may depend on the spin of the central black
hole (Meier \cite{lobanov_meier1999}). The collimation of such a jet
requires either a large scale poloidal magnetic field threading the
disk (Spruit et al. \cite{lobanov_spruit1997}) or a slower and more
massive MHD outflow launched at larger disk radii by centrifugal
forces (Bogovalov \& Tsinganos 2005).  The flowing plasma is likely to
be dominated by electron-positron pairs (Wardle et
al. \cite{lobanov_wardle1998}, Hirotani \cite{lobanov_hirotani2005})
although a dynamically significant proton component cannot be
completely ruled out at the moment (Celotti \& Fabian
\cite{lobanov_celotti1993}).

Acceleration or collimation of the flow may be complete within about
$10^3\,R_\mathrm{g}$ (Meier et al. \cite{lobanov_meier2009}) or
continue all the way to scales of a few parsecs (Vlahakis \& K\"onigl
\cite{lobanov_vlahakis2004}). At distances of
$\sim$\,$10^3\,R_\mathrm{g}$, the jets become visible in the radio
regime. Recent studies indicate that at $10^3$--$10^5\,R_\mathrm{g}$
($\le\,1$\,pc) the jets are likely to be dominated by pure
electromagnetic processes such as Poynting flux (Sikora et
al. \cite{lobanov_sikora2005}) or have both MHD (kinetic flux) and
electrodynamic components (Meier \cite{lobanov_meier2003}).  At larger
scales, the jets are believed to be kinetic flux-dominated. The
magnetic field is believed to play an important role in accelerating
and collimating extragalactic jets on parsec scales (Vlahakis \&
K\"onigl \cite{lobanov_vlahakis2004}). Three distinct regions with
different physical mechanisms dominating the observed properties of
the jets can be considered: 1)~ultra-compact jets (on scales of up to
$\sim 1$\,pc) where collimation and acceleration of the flow occurs;
2)~parsec-scale flows ($\sim$\,10\,pc scales) dominated by
relativistic shocks and 3)~larger-scale jets ($\sim$\,100\,pc) where
plasma instability gradually becomes dominant.  This picture may be
further complicated by transverse stratification of the flow, with the
jet velocity, particle density and magnetic field changing
substantially from the jet axis to its outer layers.  As a practical
result of this stratification, shocks and instabilities may in fact
co-exist on all scales in the jets, with instabilities simply
remaining undetected in compact flows, owing to limited resolution and
dynamic range of VLBI observations.

\subsection{Ultra-compact jets}

Blazar jets usually feature a bright, compact (often unresolved)
``core'' and a weaker, extended jet (often also transversely
unresolved), with several regions of enhanced emission (traditionally
branded ``jet components'') embedded in the flow and separating from
the core at apparently superluminal speeds.  The radio core is offset
from the true base of the jet, and this ``invisible'' flow is probably
Poynting flux-dominated (Meier
\cite{lobanov_meier1999,lobanov_meier2003}; Sikora et al.
\cite{lobanov_sikora2005}).

The radio core has a flat spectrum, expected to result from
synchrotron self-absorption in a conically expanding ultra-compact flow
(K\"onigl \cite{lobanov_koenigl1981}). As a result, the observed
position, $r_\mathrm{c}$, of the core depends on the frequency of
observation, $\nu$, so that $r_\mathrm{c} \propto
\nu^{-1/k_\mathrm{r}}$ (this is so called ``core shift'' effect). If
the core is self-absorbed and in equipartition, the power index
$k_\mathrm{r}=1$ (Blandford \& K\"onigl
\cite{lobanov_blandford1979}). Recent measurements (Kovalev et
al. \cite{lobanov_kovalev2008}) have also shown that the frequency
dependent core shift increases during flares as expected from the
synchrotron self-absorption.

The core shift can be used for determining basic physical properties
of the ultra-compact flow and the surrounding absorbing material
(Lobanov \cite{lobanov_lobanov1998a}).  Changes of the core position
measured between three or more frequencies can be used for determining
the value of $k_\mathrm{r}$ and estimating the strength of the
magnetic field, $B_\mathrm{core}$, in the nuclear region and the
offset, $R_\mathrm{core}$, of the observed core positions from the
true base of the jet.  The combination of $B_\mathrm{core}$ and
$R_\mathrm{core}$ gives an estimate for the mass of the central black
hole $M_\mathrm{bh} \approx 7\times 10^9\,M_\odot\,
(B_\mathrm{core}/\mathrm{G})^{1/2}
(R_\mathrm{core}/\mathrm{pc})^{3/2}$.

Core shift measurements provide estimates of the total (kinetic +
magnetic field) power, the synchrotron luminosity, and the maximum
brightness temperature, $T_\mathrm{b,max}$ in the jets can be
made. The ratio of particle energy to magnetic field energy can also
be estimated, from the derived $T_\mathrm{b,max}$.  This enables
testing the original K\"onigl model and several of its later
modifications (e.g., Hutter \& Mufson \cite{lobanov_hutter1986}; Bloom
\& Marscher \cite{lobanov_bloom1996}). The known distance from the
nucleus to the jet origin can also enable constraining the
self-similar jet model (Marscher \cite{lobanov_marscher1995}) and the
particle-cascade model (Blandford \& Levinson \cite{lobanov_blandford1995}).

Studies of free-free absorption in AGN indicate the presence of dense,
ionised circumnuclear material with $T_\mathrm{e} \approx 10^4$\,K
distributed within a fraction of a parsec of the central nucleus
(Lobanov \cite{lobanov_lobanov1998a}, Walker et
al. \cite{lobanov_walker2000}).  Properties of the circumnuclear
material can also be studied using the variability of the power index
$k_\mathrm{r}$ with frequency. This variability results from pressure
and density gradients or absorption in the surrounding medium most
likely associated with the broad-line region (BLR). Changes of
$k_\mathrm{r}$ with frequency, if measured with required precision,
can be used to estimate the size, particle density and temperature of
the absorbing material surrounding the jets. Estimates of the black hole mass and size
of the BLR obtained from the core shift measurements can be compared
with the respective estimates obtained from the reverberation mapping
and applications of the $M_\mathrm{bh}$--$\sigma_\star$ relation.

The non-thermal continuum radio emission from the jet core does not
appear to have strong shocks (Lobanov \cite{lobanov_lobanov1998b}),
and its evolution and variability can be explained by smooth changes
in particle density of the flowing plasma, associated with the nuclear
flares in the central engine (Lobanov \& Zensus
\cite{lobanov_lobanov1999}). Compelling evidence exists for
acceleration (Vlahakis \& K\"onigl \cite{lobanov_vlahakis2004}, Bach
et al. \cite{lobanov_bach2005}, Lee et al. \cite{lobanov_lee2008}) and
collimation (Junor et al. \cite{lobanov_junor1999}, Krichbaum et
al. \cite{lobanov_krichbaum2008}) in the ultra-compact flows.

The brightness temperature of the radio emission from the cores
reaches the inverse-Compton limit of $\approx 10^{12}$\,K (Unwin et
al. \cite{lobanov_unwin1997}, Lobanov et
al. \cite{lobanov_lobanov2000}, Kovalev et
al. \cite{lobanov_kovalev2005}), while it drops rapidly to the
equipartition limit of $\approx 5\times 10^{10}$\,K in the jet
components moving downstream from the core (L\"ahteenm\"aki et
al. \cite{lobanov_lahteenmaki}, Lobanov et
al. \cite{lobanov_lobanov2000}, Homan et
al. \cite{lobanov_homan2006}). This supports earlier conclusions that
ultra-compact jets are particle-dominated, while the plasma in moving
jet components is likely to be close to the equipartition (Unwin et
al. \cite{lobanov_unwin1997}, Lobanov \cite{lobanov_lobanov1998a},
Hirotani). Combining these calculations with estimates of the jet
kinetic power provides strong indications that the relativistic
fraction of the outflowing material is most likely represented by the
electron-positron plasma (Reynolds et al. \cite{lobanov_reynolds1996},
Hirotani et al. \cite{lobanov_hirotani2000}, Hirotani
\cite{lobanov_hirotani2005}).

\subsection{Parsec-scale flows: shocks and instabilities}

Parsec-scale outflows are characterised by pronounced curvature of
trajectories of superluminal components (Kellermann
\cite{lobanov_kellermann2004}, Lobanov \& Zensus
\cite{lobanov_lobanov1999}), rapid changes of velocity and flux
density (Lister et al. \cite{lobanov_lister2009}) and predominantly
transverse magnetic field~\cite{lobanov_jorstad2005}.  Statistical
studies of speed and brightness temperature distributions observed in
the superluminal features propagating on parsec scales indicate that
the jet population has an envelope Lorentz factor of $\approx 30$ and
an un-beamed luminosity of $\sim 1\times 10^{25}$\,W\,Hz$^{-1}$~(Cohen
et al. \cite{lobanov_cohen2007}).

Physical conditions of the jet plasma can be assessed effectively by
studying the spectral peak (turnover point) of the synchrotron
emission.  Mapping the turnover frequency distribution provides a
sensitive diagnostic of the plasma (Lobanov
\cite{lobanov_lobanov1998b}).  Obviously, the observed morphology and
velocity of these flows are affected substantially by Doppler
boosting, aberration, and time delays, which makes uncovering true
physical properties a non-trivial task (Gomez et
al. \cite{lobanov_gomez1994}, Cohen et al. \cite{lobanov_cohen2007}).
As a result, it becomes difficult to distinguish between apparent and
true physical accelerations of the moving features (Lister et
al. \cite{lobanov_lister2009}) and making such a distinction often
requires a detailed physical modelling of a given jet component
(Lobanov \& Zensus \cite{lobanov_lobanov1999}, Homan et
al. \cite{lobanov_homan2003}).  Similarly to stellar jets, rotation of
the flow is expected to be important for extragalactic jets (Fendt
\cite{lobanov_fendt1997}), but observational evidence remains very
limited on this issue.

VLBI studies have demonstrated that relativistic shocks are prominent
in jets on parsec scales, which is manifested by strong polarization
(Ros et al. \cite{lobanov_ros2000}) and rapid evolution of the
turnover frequency of synchrotron emission (Lobanov et
al. \cite{lobanov_lobanov1997}, Lobanov \& Zensus
\cite{lobanov_lobanov1999}).  Evidence is growing for the presence of
stationary features in parsec-scale flows, typically separated by
$\sim$\,1\,pc distance from the jet core (Kellermann et
al. \cite{lobanov_kellermann2004}, Savolainen et
al. \cite{lobanov_savolainen2006}, Lister et
al. \cite{lobanov_lister2009}, Arshakian et
al. \cite{lobanov_arshakian2010}, Le\'on-Tavares et
al. \cite{lobanov_leon2010}).  

Specific geometric conditions and extremely small viewing angles could
lead to formation of stationary features in relativistic flows (Alberdi et
al. \cite{lobanov_alberdi2000}).  However, a more general and
physically plausible explanation is offered by standing shocks (for
instance, recollimation shocks in an initially over-pressurised
outflow; Daly \& Marscher \cite{lobanov_daly1988}, G\'omez et
al. \cite{lobanov_gomez1995}, Perucho \& Mart\'i
\cite{lobanov_perucho2007a}). Such standing shocks may play a major
role in accelerating particles near the base of the jet (Mandal \&
Chakrabarti \cite{lobanov_mandal2008}, Becker et
al. \cite{lobanov_becker2008}), and could be responsible for the
persistent high levels of polarization in blazars (D'Arcangelo et
al. \cite{lobanov_darcangelo2007}, Marscher et
al. \cite{lobanov_marscher2008}).  More speculatively, the stationary
features in jets could also be the sites of continuum emission release
due to conversion from Poynting flux-dominated to kinetic
flux-dominated flow.

Complex evolution of the moving shocked regions is revealed in
observations (Gomez et al. \cite{lobanov_gomez2001}, Jorstad et
al. \cite{lobanov_jorstad2005}, Lobanov \& Zensus
\cite{lobanov_lobanov1999}) and numerical simulations (Agudo et al.
\cite{lobanov_agudo2001}) of parsec-scale outflows. However, the
shocks are shown to dissipate rapidly (Lobanov \& Zensus
\cite{lobanov_lobanov1999}), and shock dominated regions are not
likely to extend beyond $\sim$\,100\,pc. Starting from these scales,
instabilities (most importantly, Kelvin-Helmholtz instability, {\em
cf.,} Hardee \cite{lobanov_hardee2000}) determine at large the
observed structure and dynamics of extragalactic jets (Lobanov et
al. \cite{lobanov_lobanov1998c}, Walker et
al. \cite{lobanov_walker2001}, Lobanov \& Zensus
\cite{lobanov_lobanov2001}, Lobanov et al. \cite{lobanov_lobanov2003},
Hardee et al. \cite{lobanov_hardee2005}, Perucho et
al. \cite{lobanov_perucho2006}).  

The elliptical mode of the instability is responsible for appearance
of thread-like features in the jet interior , while overall
oscillations of the jet ridge line are caused by the helical surface
mode (Lobanov \& Zensus \cite{lobanov_lobanov2001}). Successful
attempts have been made to represent the observed brightness
distribution of radio emission on these scales, using linear
perturbation theory of Kelvin-Helmholtz instability~(Lobanov \& Zensus
\cite{lobanov_lobanov2001}, Lobanov et al. \cite{lobanov_lobanov2003},
Hardee et al. \cite{lobanov_hardee2005}) and a spine-sheath
(analogous to the two-fluid) description of relativistic flows (Canvin
et al. \cite{lobanov_canvin2005}, Laing \& Bridle \cite{lobanov_laing2004}).
Non-linear evolution of the instability (Perucho et
al. \cite{lobanov_perucho2004a,lobanov_perucho2004b}), stratification
of the flow (Perucho et al. \cite{lobanov_perucho2005}), and
stabilisation of the flow via magnetic field (Hardee
\cite{lobanov_hardee2007}) are important for reproducing the observed
properties of jets.  At larger scales, the helical surface mode of
Kelvin-Helmholtz instability is likely to be one of the most important
factor for disrupting and destroying the outflows (Lobanov et
al. \cite{lobanov_lobanov2003,lobanov_lobanov2006}, Perucho \& Lobanov
\cite{lobanov_perucho2007}).

\subsection{Structure of the magnetic field}

The structure of magnetic field in blazar jets can be assessed with
VLBI via linear (e.g., Lister \& Homan
\cite{lobanov_lister2005}) and circular polarisation (Homan \& Lister
\cite{lobanov_homan2006b}) measurements. The ultra-compact jets (VLBI
cores) are shown to be typically less than 5\,\% linearly polarised,
with the polarisation angle in BL Lac objects showing a stronger
tendency to be aligned with the inner jet direction (Lister \& Homan
\cite{lobanov_lister2005}). The low degree of polarisation in the VLBI
cores can be caused by a disordered magnetic field (Hughes
\cite{lobanov_hughes2005}) or strong Faraday de-polarisation (Zavala \&
Taylor \cite{lobanov_zavala2004}), or result from the ``beam
de-polarisation'' if the magnetic field in the cores is structured on
scales much smaller than the resolution of the VLBI experiments.

The fractional linear polarisation in moving jet components grows with
increasing distance from the core (Lister \& Homan
\cite{lobanov_lister2005}), and the position angle of the polarisation
vector is again better aligned with the jet direction in the BL Lac
objects. At the same time, the linearly polarised emission in
transversely resolved flows displays both strong rotation of the
polarisation angle near the core and remarkable edge brightening
(e.g., Ros et al. \cite{lobanov_ros2000}), with the polarisation
vectors tending to be perpendicular to the jet direction. This
indicates overall complexity and likely stratification of the magnetic
field in the jets, with the internal part (``spine'') of the flow
being dominated by a helical magnetic field further compressed by
relativistic shocks, while the external, slower moving layers
(``sheath'') of the flow could be dominated by a longitudinal magnetic
field. This conclusion is in a good agreement with the turnover
frequency distribution (Lobanov et al. \cite{lobanov_lobanov1997}),
internal structure (Lobanov \& Zensus \cite{lobanov_lobanov2001}), and
transverse velocity stratification in the jets (Perucho et
al. \cite{lobanov_perucho2006}). Here again, the effect that the flow,
and magnetic field, rotation (due to residual angular momentum
inherited from the initial disk-to-jet coupling) remains difficult to
assess. Presence of a strong toroidal or helical magnetic field either
in the spine or in the sheath has also been suggested from
observations of Faraday rotation gradients across the flow (Asada et
al. \cite{lobanov_asada2002}, Zavala \& Taylor
\cite{lobanov_zavala2005}, Attridge et
al. \cite{lobanov_attridge2005}, Gomez et
al. \cite{lobanov_gomez2008}).

Circular polarisation has been detected in a number of AGN jets (Homan
\& Lister \cite{lobanov_homan2006b}), with a typical level of
polarisation of $\le$0.5\,\%. The circular polarisation can be either
intrinsic to the synchrotron emission (Legg \& Westfold
\cite{lobanov_legg1968}, implying the presence of a strong
relativistic proton component) or result from scintillations (Macquart
\& Melrose \cite{lobanov_macquart2000}), relativistic effects in
dispersive plasma (Broderick \& Blandford
\cite{lobanov_broderick2003}) or Faraday conversion of linearly
polarised synchrotron emission from electron-positron plasma (Jones \&
O'Dell \cite{lobanov_jones1977}). The observed properties of
circularly polarised emission in blazar jets support the last
mechanism for its formation (Wardle et al. \cite{lobanov_wardle1998},
Homan \& Lister \cite{lobanov_homan2006b}, Homan et
al. \cite{lobanov_homan2009}).

The strength of the magnetic field in blazar jets is typically assessed by
combining multi-band measurements (Unwin et
al. \cite{lobanov_unwin1997}), by deriving information about the peak
in the synchrotron spectrum (Marscher \cite{lobanov_marscher1983},
Lobanov et al. \cite{lobanov_lobanov1997}, Lobanov
\cite{lobanov_lobanov1998b}, Savolainen et
al. \cite{lobanov_savolainen2008}, Sokolovsky et
al. \cite{lobanov_sokolovsky2010}), or by using the opacity due to
synchrotron self-absorption (Lobanov \cite{lobanov_lobanov1998a}). In most objects, values around 1\,G are obtained for the VLBI cores and lower magnetic field is measured in the jets, all falling well in agreement with magnetic field generation being ultimately processes in the magnetised accretion disk (Field \& Rogers \cite{lobanov_field1993}).

\subsection{Periodic changes of the structure}

Structural changes are abound on milliarcsecond scales in the blazar
jets, enhanced and magnified by small viewing angles and relativistic
effects. In addition to extreme curvature observed in the jet ridge
line of some objects (e.g., Polatidis et
al. \cite{lobanov_polatidis1995}), the position angle of the inner jet
(as traced by jet components nearest to the core) changes
substantially both as a function of observing frequency (Savolainen et
al. \cite{lobanov_savolainen2006}, Agudo et
al. \cite{lobanov_agudo2007}) and in time (Mutel \& Denn 
\cite{lobanov_mutel2005}, Lobanov \& Roland
\cite{lobanov_lobanov2005}). 

The frequency dependent changes are most likely caused by the opacity
and spectral index gradients in the flow. The temporal variations of
the position angle can result from precession and rotation of the flow
(Camenzind et al. \cite{lobanov_camenzind1992}) as well as from the
pattern motion of Kelvin-Helmholtz instability (Hardee
\cite{lobanov_hardee2003}, Hardee et al. \cite{lobanov_hardee2005}),
in which case the motion should also be evident in long-term evolution
of the ridge line of a flow (e.g., Krichbaum et
al. \cite{lobanov_krichbaum2001}).

Variations of the jet position angle, as well as the observed
morphology of parsec-scale jets and trajectories of superluminal
features propagating in the jets, are often described in terms of a
helical geometry (Steffen et al. \cite{steffen1995}), with helicity
supposed to be arising from some periodic process in the nucleus.  Jet
precession, both in single and binary black hole systems, have been
commonly sought to be responsible for the observed helicity on parsec
scales. However, as the observed periods of the position angle changes
(and, similarly, periods inferred from the component trajectories and
oscillations of the jet ridge lines) are typically within a range of a
few years, the precession-based models face severe difficulties
(Lobanov \& Roland \cite{lobanov_lobanov2005}) as they require either
allowing for extremely small ($\le 10^4\,R_\mathrm{g}$) orbital
separations in supermassive binary black holes or adopting an
assumption that the jet direction responds exclusively to changes in
the innermost parts of the accretion disk and it is decoupled from the
spin of the central black hole. In view of these difficulties,
rotation of the flow and pattern motion of the instability seem to be
more viable alternatives.  The precession of the flows, evident on
kiloparsec-scales (Gower et al. \cite{lobanov_gower1982}, Hardee et
al. \cite{lobanov_hardee1994}) acts on much longer timescales
(typically $\ge 10^4$\,yr), typically of several hundred years and
longer, and thus should only be visible in long-term changes of the
position angle of the entire milliarcsecond-scale jet (Lobanov \&
Roland \cite{lobanov_lobanov2005}).

\section{Emission from blazar jets}

VLBI observations enable tracing both temporal and spatial changes in
radio emission from blazar jets, offering a unique opportunity for
connecting these changes to properties of the blazar emission observed
in other wavelength domains and even localising spatially the dominant
components in the broad-band emission. Emission from jets may also
contribute substantially to the broad-band SED (Yuan et
al. \cite{lobanov_yuan2002}), alongside with the canonical
contributions from the accretion disk and hot material in the vicinity
of the central black hole. VLBI observations help here enormously by
providing an accurate measure of radio emission produced in these
regions and excluding contributions from kiloparsec-scale jets and
lobes of radio sources.

\subsection{Parsec-scale radio emission}

Parsec-scale radio emission is variable on timescales from decades
(e.g., O'Dea et al. \cite{lobanov_odea1984}, Asada et
al. \cite{lobanov_asada2006}) to hours (Savolainen \& Kovalev
\cite{lobanov_savolainen2008}), with the longest timescales most
likely related to large-scale changes in the nuclear region feeding
the black hole, and the shortest timescales resulting from
interstellar scintillations. The variability on hour-to-day timescales
may also be related to ``quasar QPO'' type of variations recently
observed in the soft X-ray band (Gierl\'inski et
al. \cite{lobanov_gierlinski2008}), remembering that if variability
periods scale with the black hole mass, this would correspond to
canonical QPO seen in X-ray binaries at frequencies around 100\,Hz.

On intermediate timescales (months-to-years), most of variable radio
emission is believed to be associated with flares in the VLBI cores
(Lobanov \& Zensus \cite{lobanov_lobanov1999}) and shock evolution of
plasma propagating downstream (Hughes et
al. \cite{lobanov_hughes1985}, Marscher \& Gear
\cite{lobanov_marscher1985}, Marscher
\cite{lobanov_marscher1990}). The radio flares last, on average, for
2.5 years (at a wavelength of about 1\,cm; Hovatta et
al. \cite{lobanov_hovatta2008}), and in many objects they are repeated
quasi-periodically (Hovatta et al. \cite{lobanov_hovatta2007}).  The
flares are firmly associated with ejections of new jet components
(Valtaoja et al. \cite{lobanov_valtaoja1999}, Lobanov \& Zensus
\cite{lobanov_lobanov1999}, Marscher et
al. \cite{lobanov_marscher2002}, Chatterjee et
al. \cite{lobanov_chatterjee2009}), but the release of non-thermal
continuum emission may not necessarily be restricted to the vicinity
of the black hole or even the radio core (Arshakian et
al. \cite{lobanov_arshakian2010}, Leon-Tavares et
al. \cite{lobanov_leon2010}, Schinzel et
al. \cite{lobanov_schinzel2010}).  Detailed evolution of a flaring
emission is best determined through variations of its turnover peak
(Otterbein et al. \cite{lobanov_otterbein1998}, Fromm et
al. \cite{lobanov_fromm2010}) and can be further constrained using
observed kinematic properties of an emitting region (Lobanov \& Zensus
\cite{lobanov_lobanov1999}).

Quasi-periodic variability of the radio emission from the ultra-compact
jets is most likely related to instabilities and non-stationary
processes in the accretion disks around central black holes in AGN
(Igumenschev \& Abramowicz \cite{lobanov_igumenschev1999}, Lobanov \&
Roland \cite{lobanov_lobanov2005}. Alternative explanations involve
binary black hole systems in which flares are caused by passages of
the secondary through the accretion disk around the primary Ivanov et
al. \cite{lobanov_ivanov1998}, Lehto \& Valtonen \cite{lobanov_lehto1996}. Similarly to the attempts of using binary
black holes to explain short-term morphological changes, these
models require very tight binary systems, with orbits of the secondary
lying well within $10^4\,R_\mathrm{g}$ of the primary, which
poses inevitable problems for maintaining an accretion disk around the
primary (for massive secondaries; Lobanov \cite{lobanov_lobanov2008})
or rapid alignment of the secondary with the plane of the accretion
disk (for small secondaries; Ivanov et al. \cite{lobanov_ivanov1999}).

\subsection{Blazar jets and broad-band continuum}

Relativistic flows are prominent emitters in all bands of the
electromagnetic spectrum, generating optical and X-ray emission even
on kiloparsec scales (Schwartz et al. \cite{lobanov_schwartz2000},
Marshall et al. \cite{lobanov_marshall2002}, Siemiginovska et
al. \cite{lobanov_siemiginovska2002}, Sambruna et
al. \cite{lobanov_sambruna2008}) and at TeV energies (Acciari et
al. \cite{lobanov_acciari2010}). The jet plasma is believed to emit via synchrotron emission in the radio to
soft X-ray range and via inverse Compton emission in the hard X-ray to TeV
range (Acciari et al. \cite{lobanov_acciari2010}, Marscher et
al. \cite{lobanov_marscher2010}). Contested is, however, the primary
source of the seed photons for the inverse Compton part of the
radiation. These could be the synchrotron photons themselves
(synchrotron self Compton mechanism, SSC) or photons from an external
radiation field located, for instance, in the accretion disc emission,
X-ray corona, the broad-line region, the infrared emitting torus and
the cosmic background radiation (Ghisellini \& Tavecchio
\cite{lobanov_ghisellini2009}).

Results from the first year of the \fermilat\ operations have
enabled most detailed studies of the connection between relativistic
flows and $\gamma$-ray production in blazars.  Early statistical
comparisons of the properties of $\gamma$-ray emission and compact
radio jets in blazars indicate unequivocally that they are closely
related (Pushkarev et al. \cite{lobanov_pushkarev2009}, Savolainen et
al. \cite{lobanov_savolainen2010}).  Several of the $\gamma$-ray
flares detected with \fermilat\ can be associated with emission from
accelerated plasma cloud embedded in the jets (Marscher et al
\cite{lobanov_marscher2010}, Schinzel et
al. \cite{lobanov_schinzel2010}), although detailed localisation of
the $\gamma$-ray emitting sites remain elusive, clearly calling for
continuation of extensive, co-ordinated \fermilat\ and VLBI campaigns.

There is now growing evidence for the broad-band continuum (and its
flaring components in particular) to be produced at multiple locations
in AGN (Arshakian et al. \cite{lobanov_arshakian2010}), with the
emission in different bands dominated by contributions from spatially
different regions (Leon-Tavares \cite{lobanov_leon2010}). These
findings put an additional strain on single-zone models commonly used
for fitting the broad-band SED in AGN.  The situation with the
quiescent (and slowly variable) component of the $\gamma$-ray emission
is also puzzling, with indications that it may be produced in the
region of the jet extending up to $\sim$10\,pc (Schinzel et
al. \cite{lobanov_schinzel2010}).

These results give a compelling indication that formation of jet
components may correspond to the strongest nuclear events, while some
of them may not survive a passage through a standing shock
(Leon-Tavares et al. \cite{lobanov_leon2010}). The source of the
continuum emission is localised not only in the accretion disk (at the
extreme vicinity of the black hole) but also in the entire
acceleration zone of the jet, with strong flares happening both near
the central black hole and at a standing shock in the jet.  Clearly, a
more general and systematic study of relation between the radio,
optical and X-ray emission in blazars is strongly justified, and
co-ordinated VLBI--\fermilat\ campaigns would one of the prime methods
for such studies.

\section{Outlook}

More than thirty years after their appearance on the scientific scene,
blazars remain one of the focal points of extragalactic astrophysics.
High-resolution radio observations of blazars provide essential
information about structure, kinematics and emission of relativistic
flows in these objects on scales inaccessible to direct imaging in
other bands. This information turns out to be arguably an
indispensable tool for constructing viable physical models capable of
explaining the blazar phenomenon and its spectacular observational
manifestations.  

There are still a number of unanswered
questions about the jets themselves and their relations to the
broad-band emission produced in blazars. 
Extragalactic jets are an excellent laboratory for studying physics of
relativistic outflows and probing conditions in the central regions of
active galaxies. Recent studies of extragalactic jets show that they
are formed in the immediate vicinity of central black holes in
galaxies and carry away a substantial fraction of the angular momentum
and energy stored in the accretion flow and rotation of the black
hole. The jets are most likely collimated and accelerated by
electromagnetic fields. Relativistic shocks are present in the flows
on small scales, but dissipate on scales of decaparsecs. Plasma
instabilities dominate the flows on larger scales.  Convincing
observational evidence exists, connecting ejections of material into
the flow with instabilities in the inner accretion disks. 

In the coming years, new breakthroughs should be coming in studies of
relativistic jets and blazars, enriched by \fermilat\ results and by
an effective use of the VLBI potential for studies of blazars. There
are several potential focal points for these studies.

In particular, the jet composition remains an open issue, in particular
with regard to the role of relativistic protons in high-energy
emission production in extreme vicinity of the central black holes
(while it seems that pair plasma is chiefly responsible for the
emission from parsec-scale jets).

Accurate spatial localisation of the sites of high-energy continuum
production will play a crucial role for modelling the broad-band SED
of blazars and understanding their physical nature in general. To this
end, variability of the high-energy continuum emission is best related
to structural and radiative changes in parsec-scale radio emission
resolved by VLBI observations.

Last but not least, the continued observational quest for reaching
ever closer to the regions where the relativistic flows are formed
will bring new answers not only about the physical nature of
relativistic flows, but also about the physical properties of black
holes and their connection to major manifestations of nuclear activity
in galaxies. This would be particular important for studies of the
$\gamma$-ray and TeV emission from blazars, as this emission has all
chances to come from the immediate vicinity of the central black holes
in galaxies. Combination of high-energy observations with VLBI
extensive monitoring at centimetre wavelengths and focused, specific
VLBI programs at millimetre wavelengths certainly has a quality of
the tool of choice for such studies.

\end{document}